# An Energy Activity Dataset for Smart Homes

Chen Li, *Student Member*, *IEEE*

*Abstract*—A smart home energy dataset that records miscellaneous energy consumption data is publicly offered. The proposed energy activity dataset (EAD) has a high data type diversity in contrast to existing load monitoring datasets. In EAD, a simple data point is labeled with the appliance, brand, and event information, whereas a complex data point has an extra application label. Several discoveries have been made on the energy consumption patterns of many appliances. Load curves of the appliances are measured when different events and applications are triggered and utilized. A revised longest-common-subsequence (LCS) similarity measurement algorithm is proposed to calculate energy dataset similarities. Thus, the data quality prior information becomes available before training machine learning models. In addition, a subsample convolutional neural network (SCNN) is put forward. It serves as a non-intrusive optical character recognition (OCR) approach to obtain energy data directly from monitors of power meters. The link for the EAD dataset is: https://drive.google.com/drive/folders/1zn0V6Q8eXXSKxKgcs8ZRValL5VEn3anD

## I. Introduction

Deep learning (DL) plays a vital role in smart home energy projects. State-of-the-art achievements in load identification [1], load event detection [2], and load forecasting [3] are often implemented by DL. The purpose of DL-powered energy projects lies primarily in three aspects: Firstly, a complete energy consumption profile fosters homeowners' energy-saving habits. An energy recommender system creates optimal energy usage schedules for electricity consumers. Secondly, home automation programs based on energy utilization data and internet of things (IoT) technologies improve homeowners' comfort level of living. Such programs include automatic cooling, heating, lighting adjustment, and home security surveillance. Thirdly, as an increasing number of intelligent devices enter millions of households, privacy and security protections have become crucial in smart home research [4]. DL-powered energy projects can monitor appliance usage data and detect malicious attacks in time, *e.g.*, illegal cryptocurrency mining [5].

DL-powered energy projects heavily rely on energy datasets, *i.e.*, a neural network without training data is analogous to an engine without fuel. With the evolution of hardware acceleration technologies, deeper neural networks are designed to tackle challenging load identification, disaggregation, and event detection tasks. However, complex DL models require more training data to resolve overfitting issues. Generated artificial data may not represent the real distribution, *e.g.*, although random flipping and rotation are viable data augmentation techniques for image classification tasks, reversing time-series-based energy data is impractical. Thus, collecting authentic and labeled energy data is crucial for training complex DL models. The DL models will become increasingly accurate as new samples are obtained from the real distribution during the cumulative data collection process.

Although public energy datasets are available [6], [7], [8], [9], [10], [11], several problems exist in these datasets. First, the number of energy datasets is significantly less than the dataset quantity in other research disciplines, such as computer vision and natural language processing. Second, the data type diversity of existing energy datasets is low in that the appliance, brand, event, and application/software labels are unavailable. For instance, there are only eleven types of appliances in the PLAID dataset [10], six of which are conspicuously distinguishable. The COOLL dataset [9] only records the on-off events. Third, the low sampling frequency makes capturing instant energy events almost infeasible, especially for power-on and power-off events that contain ample appliance identity information. Part of the UK-DALE [11] data is collected by 1Hz meters, which makes load identification tasks challenging. Fourth, physical quantities are incomplete in some energy datasets. For example, the PLAID dataset offers only the voltage and current data. However, voltage, current, apparent power, active power, and power factor quantities should be provided to depict an energy event comprehensively. Other energy dataset issues include: 1) update discontinuation; 2) reluctance to disclose user behavior data; 3) only the aggregated energy data are available; iv) inaccuracy; 4) disproportional data points in terms of category.

Traditional ways to collect energy data are inflexible and cumbersome. For power meters without network functionalities, exporting energy data to workstations requires complicated hardware configurations and embedded programming expertise. Unfortunately, the programs for hardware-level data retrieval are not platform-independent, *e.g.*, programs may require revision when the power meter is replaced. For power meters with network functionalities, heterogeneous network communication protocols make power meter replacement and upgrade difficult. The inconvenient energy data collection procedure undermines the volume and diversity of energy datasets. Therefore, a non-intrusive solution is required to collect energy data.

This paper proposes a novel method for creating an energy activity dataset. Detailed labels, including appliance type, appliance brands, applications of appliances (the application label is available if the appliance has applications or software installed, *e.g.*, cell phone or laptop), and energy usage events, are prepared to analyze energy consumption patterns. With all four labels combined, an "energy activity" describes a homeowner's behavior of using an appliance of a particular brand to perform a task, possibly through an application or software of the appliance. Besides, sub-datasets for different DL tasks can be easily created using permutation. Theoretically, $2^4-1$ types of sub-datasets can be created, given that the application label exists. Otherwise, $2^3-1$ kinds of sub-datasets can be created. The EAD dataset contains plenty of data points for low-powered appliances so that homeowner behaviors are better described. A revised LCS similarity measurement algorithm is proposed to calculate data point similarities of an energy dataset. This algorithm offers data quality prior information before training machine learning models.

A non-intrusive OCR-based energy data collection procedure is proposed, which allows for collecting data directly from the power meter monitor. Contrary to prior works on OCR-based meter reading that rely on conventional CNNs [12], [13], [14], the SCNN is designed to improve digit image recognition accuracy. An automatic correction mechanism is also proposed to fine-tune energy data according to physics rules. The rules formulate constraints for voltage, current, active power, apparent power, and power factor quantities.

This paper makes the following contributions:
(1) An energy activity dataset with detailed labeling and high data type diversity is publicly provided. Energy consumption patterns of applications and events are measured based on appliances from various brands.
(2) A revised LCS similarity measurement algorithm is proposed to provide prior information on energy dataset quality before training machine learning models.
(3) A non-intrusive OCR-based energy data collection procedure is put forward based on the proposed SCNN and the physics quantity auto-correction mechanism.

This paper is organized as follows: Section II illustrates the creation of the energy activity dataset. Section III describes the energy data collection procedure. Section IV summarizes the paper.

## II. ENERGY ACTIVITY DATASET CREATION

The energy activity dataset provides training data for DL-based energy projects. The dataset can be applied to energy conservation tasks such as energy usage profile creation and identification of energy-consuming appliances. The dataset can also be utilized by home automation applications to analyze homeowner behaviors. DL-based appliance attack detection models can be trained using the dataset because malicious activities impact energy consumption. This section offers an overview of the energy activity dataset, proposes an energy dataset similarity measurement algorithm based on LCS, and discusses energy consumption patterns of complex and simple appliances.

### A. Energy Activity Dataset

Six physical quantities are provided in EAD to describe energy activities: voltage ($u$), current ($i$), apparent power ($s$), active power ($p$), power factor ($\cos\varphi$), and frequency ($f$). $i$, $s$, $p$, and $\cos\varphi$ are directly influenced by energy activities, while $u$ and $f$ are indicators of grid stability. Each data point stores all six quantities within a short duration to describe an energy activity. The OCR sampling frequency is 5 Hz, *i.e.*, five frames are captured by the camera in a second. Fig. 1 illustrates a data point in EAD. For the rest of the paper, "power-on" and "open application" events are denoted by purple triangle symbols; "power-off" and "close application" events are denoted by purple square symbols; "level increment" and "level decrement" events are denoted by yellow triangle and square symbols; self-defined starting and ending events are denoted by lake blue triangle and square symbols.

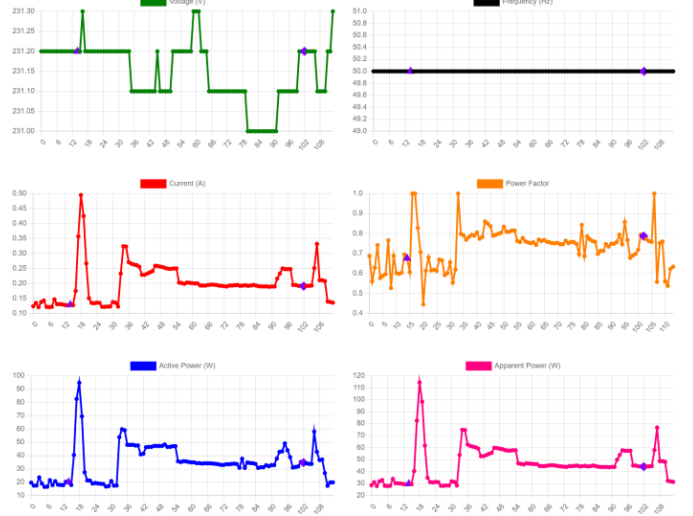

Fig. 1. An energy activity of opening and closing the Chrome browser using a Dell G3 3590 laptop computer.

EAD has two labeling systems: appliance-brand-application-event and appliance-brand-event. Applications can be installed on a **complex appliance**, *e.g.*, computers and smartphones, while a **simple appliance** has no application, *e.g.*, fans and heaters. Thus, the first labeling system is for complex appliances, and the second labeling system is for simple appliances. The appliance label represents a set of devices with similar functionalities, *e.g.*, a refrigerator or television. The brand label combines the manufacturer name and the type of an appliance, *e.g.*, Dell-Inspiron-15-7547. The application label denotes the software installed on an appliance, *e.g.*, YouTube. The event label corresponds to an activity associated with human actions or automated programs, *e.g.*, powering on a laptop or scheduled launch of anti-virus software. An **energy activity** is defined as a person triggering an event using an appliance of a particular brand, possibly through an application.

### B. Energy Dataset Similarity Measure

Solving DL-based load identification and load event detection problems requires high-quality energy datasets. The similarity is a quality indicator of energy datasets. In supervised learning, **homogeneous** data points come from the same distribution, whereas **heterogeneous** data points originate from different distributions. In EAD, labels of all homogeneous data points are the same, while at least one label differs for two heterogeneous data points. Consequently, homogeneous data points are expected to have a higher similarity than heterogeneous data points. A dataset containing homogeneous data points has limited research value if its similarity is too low. Likewise, a dataset with heterogeneous data points has little research significance if its similarity is too high. Thus, an effective similarity measurement approach is essential for evaluating energy dataset qualities. This paper proposes a two-step approach to measure energy dataset similarity: 1) measure the similarity of two time series; 2) create similarity matrices.

In the first step, a revised LCS approach is used for measuring the similarity between two time series. Although Euclidian distance and dynamic programming (DP) are two common measures for this task, the euclidian methods are sensitive to signal shifts and not robust to noise and outliers [15]. The dynamic time wrapping (DTW) method is not robust to noisy data [16]. This paper proposes a revised LCS similarity measurement algorithm based on [16]. Define $x_i$ as the $i^{th}$ scalar of time series $\mathbf{x}$, $\mathbf{x}_i = (x_1, x_2, \ldots, x_i)$; $y_j$ as the $j^{th}$ scalar of time series $\mathbf{y}$, $\mathbf{y}_j = (y_1, y_2, \ldots, y_j)$; len as a function that obtains the length of a time series; lcs as a function that obtains the length of the longest common subsequence of two time series; sim as the similarity function; usm as the similarity upper bound function. Since the original LCS algorithm only handles discrete data, the condition that two continuous scalars are equal should be redefined in a fuzzy manner:

$$|x_i - y_j| < \varepsilon \quad (1)$$

However, the similarity measurement is subjectively affected by the manually-tunned hyperparameter $\varepsilon$. To resolve this issue, $\varepsilon$ is defined as the minimal standard deviation of $\mathbf{x}$ and $\mathbf{y}$. The state-transition equations are adjusted as:

$$\text{lcs}(\mathbf{x}_i, \mathbf{y}_j) = \begin{cases} \text{lcs}(\mathbf{x}_{i-1}, \mathbf{y}_{j-1}) + 1, & |x_i - y_j| < \varepsilon \\ \max\{\text{lcs}(\mathbf{x}_i, \mathbf{y}_{j-1}), \text{lcs}(\mathbf{x}_{i-1}, \mathbf{y}_j)\}, & |x_i - y_j| \geq \varepsilon \end{cases} \quad (2)$$

$$\varepsilon = \min\{\text{std}(\mathbf{x}), \text{std}(\mathbf{y})\}$$

The similarity between $\mathbf{x}$ and $\mathbf{y}$ is:

$$\text{sim}(\mathbf{x}, \mathbf{y}) = \frac{2\text{lcs}(\mathbf{x}, \mathbf{y})}{\text{len}(\mathbf{x}) + \text{len}(\mathbf{y})} \quad (3)$$

The lower and upper bounds of $\text{sim}(\mathbf{x}, \mathbf{y})$ are:

$$\text{sim}(\mathbf{x}, \mathbf{y}) \in \left[\frac{\text{lcs}(\mathbf{x}, \mathbf{y})}{\max\{\text{len}(\mathbf{x}), \text{len}(\mathbf{y})\}}, \frac{\text{lcs}(\mathbf{x}, \mathbf{y})}{\min\{\text{len}(\mathbf{x}), \text{len}(\mathbf{y})\}}\right] \quad (4)$$

Although $\text{sim}(\mathbf{x}, \mathbf{y})$ is the de-facto LCS similarity, the upper bound of $\text{sim}(\mathbf{x}, \mathbf{y})$ denoted by $\text{usm}(\mathbf{x}, \mathbf{y})$ is more appropriate for measuring energy dataset similarities because usm is less sensitive to time series length difference, i.e., even if $\text{len}(\mathbf{x}) \ll \text{len}(\mathbf{y})$, as long as $\mathbf{x}$ is the sub-time-series of $\mathbf{y}$, then $\text{usm}(\mathbf{x}, \mathbf{y}) = 1$.

In the second step, the self-similarity matrix and cross-similarity matrix are created. In the first case, let $N$ be the number of homogeneous data points in dataset $D$; $\mathbf{x}^{(i)}$ be the $i^{th}$ data point in $D$; The $N \times N$ symmetric self-similarity matrix $\mathbf{S}(N)$ is defined as:

$$\begin{pmatrix} \text{usm}(\mathbf{x}^{(1)}, \mathbf{x}^{(1)}) & \text{usm}(\mathbf{x}^{(1)}, \mathbf{x}^{(2)}) & \cdots & \text{usm}(\mathbf{x}^{(1)}, \mathbf{x}^{(N)}) \\ \text{usm}(\mathbf{x}^{(2)}, \mathbf{x}^{(1)}) & \text{usm}(\mathbf{x}^{(2)}, \mathbf{x}^{(2)}) & \cdots & \text{usm}(\mathbf{x}^{(2)}, \mathbf{x}^{(N)}) \\ \vdots & \vdots & \ddots & \vdots \\ \text{usm}(\mathbf{x}^{(N)}, \mathbf{x}^{(1)}) & \text{usm}(\mathbf{x}^{(N)}, \mathbf{x}^{(2)}) & \cdots & \text{usm}(\mathbf{x}^{(N)}, \mathbf{x}^{(N)}) \end{pmatrix} \quad (5)$$

The similarity of the dataset $D$ is:

$$\text{usm}(D) = N^{-2} \|\mathbf{S}(N)\|_1 \quad (6)$$

The quality of $D$ is unsatisfactory when $\text{usm}(D)$ is too low. In the second case, the dataset $D$ consists of several sub-datasets: $D = D_1 \cup D_2 \cup \cdots \cup D_\Omega$. Each sub-dataset contains homogenous data points, where labels of all homogenous data points are the same. Let $D_a$ and $D_b$ be any of the two sub-datasets of $D$; $M$ and $N$ be the number of data points in $D_a$ and $D_b$; $\mathbf{x}^{(i)}$ be the $i^{th}$ data point in $D_a$; $\mathbf{y}^{(j)}$ be the $j^{th}$ data point in $D_b$. The $M \times N$ cross-similarity matrix $\mathbf{S}(M, N)$ is defined as:

$$\begin{pmatrix} \text{usm}(\mathbf{x}^{(1)}, \mathbf{y}^{(1)}) & \text{usm}(\mathbf{x}^{(1)}, \mathbf{y}^{(2)}) & \cdots & \text{usm}(\mathbf{x}^{(1)}, \mathbf{y}^{(N)}) \\ \text{usm}(\mathbf{x}^{(2)}, \mathbf{y}^{(1)}) & \text{usm}(\mathbf{x}^{(2)}, \mathbf{y}^{(2)}) & \cdots & \text{usm}(\mathbf{x}^{(2)}, \mathbf{y}^{(N)}) \\ \vdots & \vdots & \ddots & \vdots \\ \text{usm}(\mathbf{x}^{(M)}, \mathbf{y}^{(1)}) & \text{usm}(\mathbf{x}^{(M)}, \mathbf{y}^{(2)}) & \cdots & \text{usm}(\mathbf{x}^{(M)}, \mathbf{y}^{(N)}) \end{pmatrix} \quad (7)$$

The similarity between $D_a$ and $D_b$ is:

$$\text{usm}(D_a, D_b) = (MN)^{-1} \|\mathbf{S}(M, N)\|_1 \quad (8)$$

The qualities of $D_a$ and $D_b$ are unsatisfactory when $\text{usm}(D_a, D_b)$ is too high. The similarity of the dataset $D$ is a $C_\Omega^2$ dimensional vector:

$$\mathbf{usm}(D) = (\text{usm}(D_1, D_2), \text{usm}(D_1, D_3), \ldots, \text{usm}(D_{\Omega-1}, D_\Omega)) \quad (9)$$

### C. Complex Appliance Patterns

Four scenarios are discussed for the complex appliances according to Table I:

Table I Energy Activity Scenarios for Complex Appliances

| Scenario | Appliance | Brand | Application | Event |
|---|---|---|---|---|
| I | Same | Same | Same | Same |
| II | Same | Same | Different | Same |
| III | Same | Different | Same | Same |
| IV | Same | Same | Same | Different |

In the first scenario, identical measurements are conducted independently. The appliance, brand, application, and event labels are the same for each measurement. According to Fig. 2, highly similar energy consumption patterns are observable. **Table III** provides the dataset description and self-similarities for complex appliances.

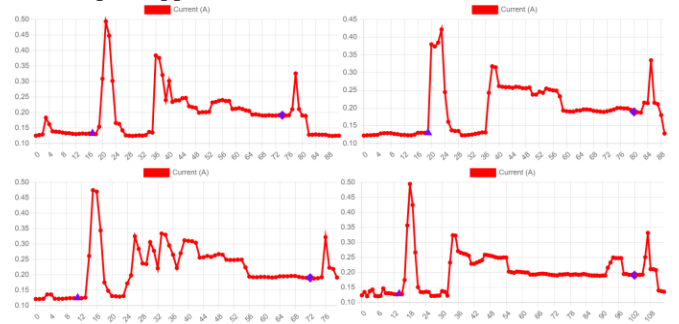

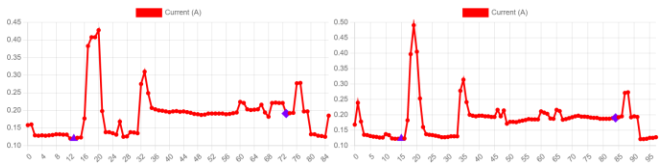

Fig. 2. Six current time series depict energy activities of opening and closing the Chrome browser using the Dell-G3-3590 laptop.

In the second scenario, different applications are launched. The appliance, brand, and event labels are the same for each measurement. Fig. 3 shows that different applications have conspicuous distinctions in energy consumption patterns. Nevertheless, applications from the same product family share similar patterns, *e.g.,* Microsoft Office applications.

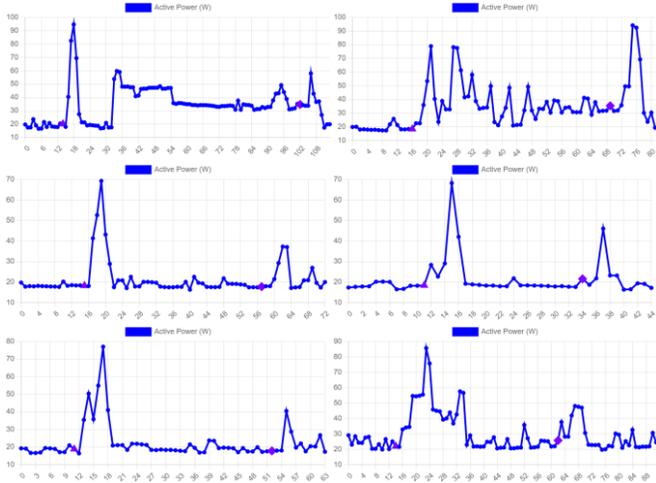

Fig. 3. Six active power time series depict energy activities of opening and closing the Chrome, Edge, Excel, OneNote, PowerPoint, and Word applications using a Dell-G3-3590 laptop.

Fig. 4 offers an example of the cross-similarity bar chart in the second scenario, *e.g.,* the finance application AliPay and the e-commerce application JD are only 23% similar, making them highly distinguishable.

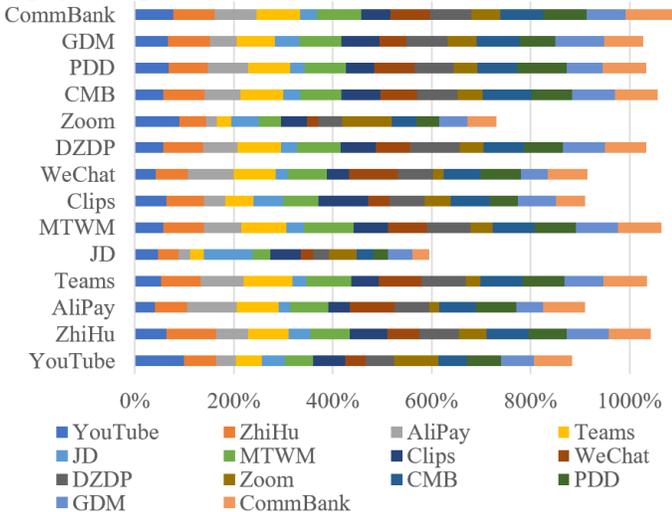

Fig. 4. Cross-similarities for opening and closing applications using an iPhone-10. Full Chinese names of the application acronyms are 高德地图(GDM), 拼多多(PDD), 招商银行(CMB), 大众点评(DZDP), 美团外卖(MTWM).

In the third scenario, the same application is run on appliances from different brands. The appliance, application, and event labels are the same for each measurement. Fig. 5 shows that hardware differences significantly impact energy consumption patterns. However, similar features in terms of peaks and valleys are still noticeable.

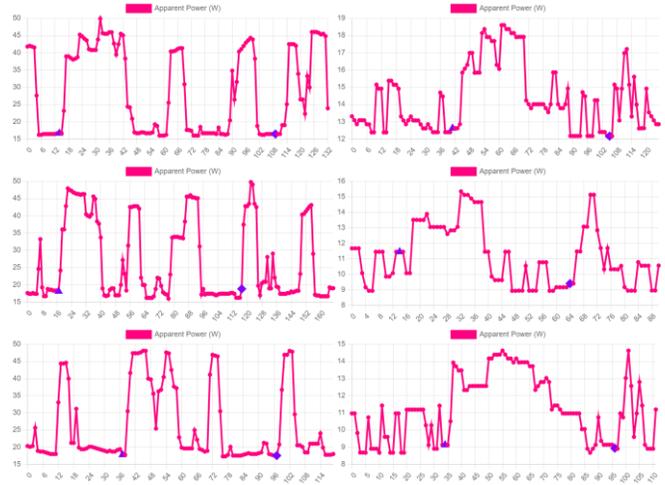

Fig. 5. Three apparent power time series on the left and right columns depict energy activities of opening and closing the Word, Excel, and PowerPoint applications using a Dell-Inspiron-15-7547 laptop computer and a Surface-1 tablet computer respectively.

Fig. 6 offers an example of the cross-similarity bar chart in the third scenario. All three devices are highly distinguishable.

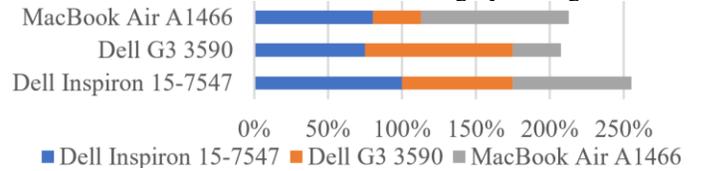

Fig. 6. Cross-similarities for opening and closing the excel application using different laptops.

In the fourth scenario, different events are triggered. The appliance, brand, and application labels are the same for each measurement. Fig. 7 shows that each event can have a distinctive energy consumption pattern.

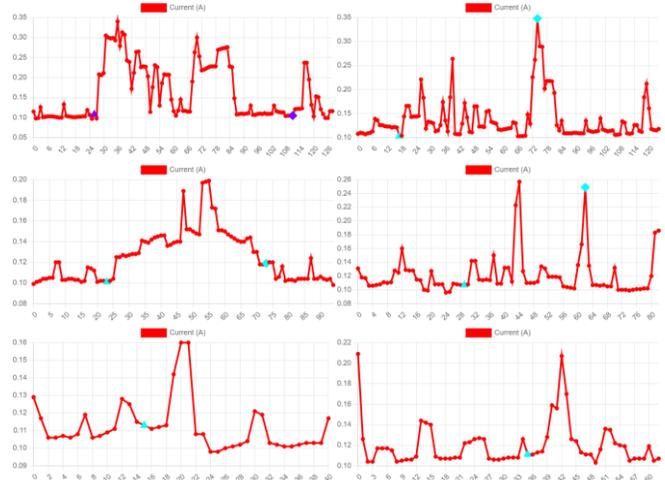

Fig. 7. Six current time series depict energy activities when the Word application is used on a Dell-G3-3590 laptop : opening and closing a document, document saving, typing, image insertion, word count, and paragraph navigation.

Fig. 8 offers an example of the cross-similarity bar chart in the fourth scenario. Events such as playing videos and scanning

QR codes are highly distinguishable because they consume more energy than other events.

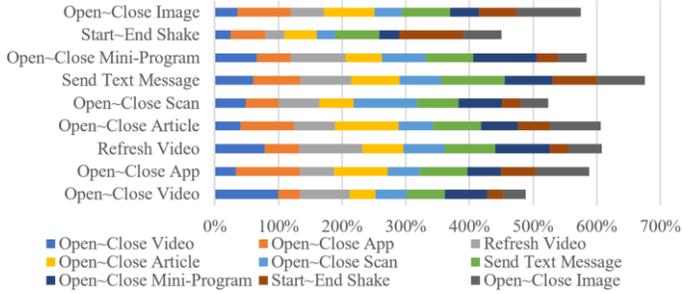

Fig. 8. Cross-similarities for triggering different events using the WeChat application on iPhone-10.

*D. Simple Appliance Patterns*

Three scenarios are discussed for simple appliances according to Table II:

Table II Scenarios for Simple Appliances

| Scenario | Appliance | Brand | Event |
|---|---|---|---|
| I | Same | Same | Same |
| II | Different | Different | Same |
| III | Same | Same | Different |

In the first scenario, identical measurements are conducted independently. The appliance, brand, and event labels are the same for each measurement. Fig. 9 shows that common energy consumption patterns are conspicuously observed. **Table IV** offers the dataset description and self-similarities in the first scenario.

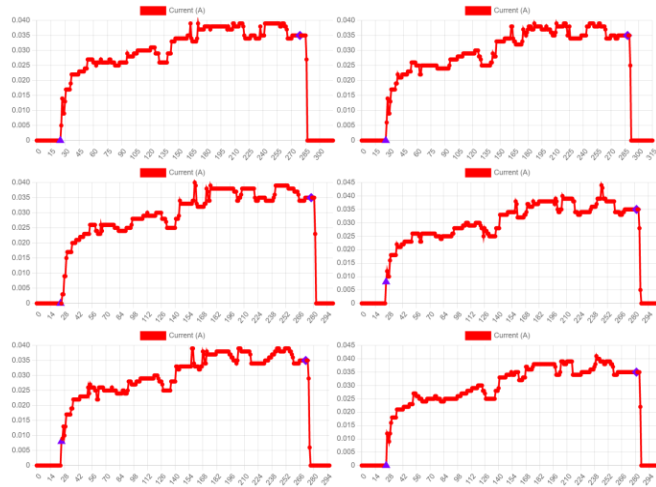

Fig. 9. Six current time series depict energy activities of powering on and off a XiaoMi-4A-1200M router.

In the second scenario, the same events are triggered on different appliances. Fig. 10 shows that the active power magnitude can be a significant indicator of appliance type [1].

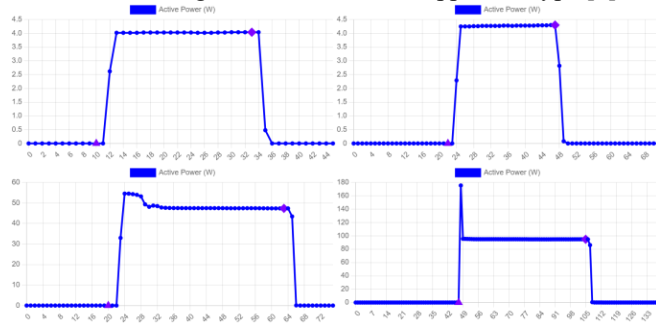

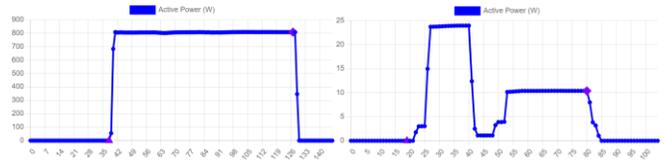

Fig. 10. Six active power time series depict energy activities of powering on and off the fluorescent lamp, incandescent lamp, LED lamp, massager gun, heater, and monitor appliances.

Fig. 11 offers an example of the cross-similarity bar chart in the second scenario. Appliances with special event programs are easily distinguishable.

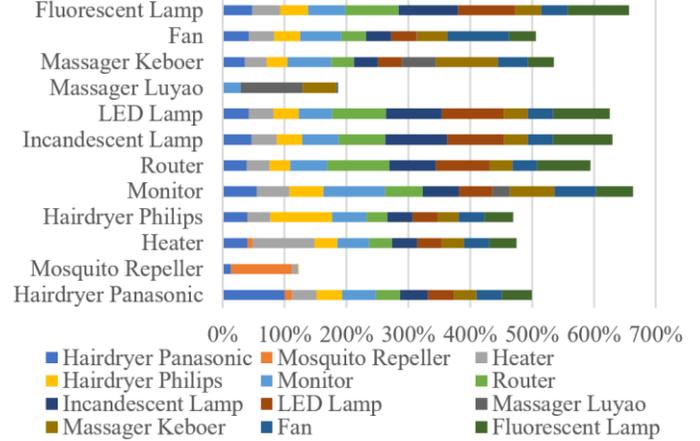

Fig. 11. Cross-similarities for different appliances when on-to-off events are triggered.

In the third scenario, different events are triggered. The appliance and brand labels are the same for each measurement. Fig. 12 shows that energy consumption patterns vary under different events.

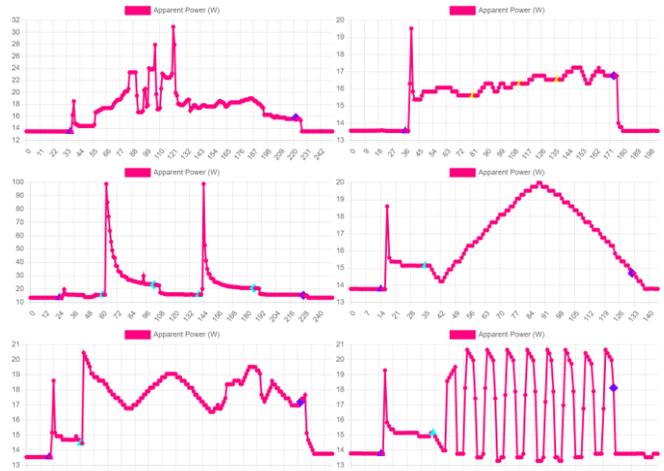

Fig. 12. Six apparent power time series depict energy activities when using a massager gun: power-on-to-power-off, intensity increment, heating, relax mode, refresh mode, and excite mode.

Fig. 13 offers an example of the cross-similarity bar chart in the third scenario.

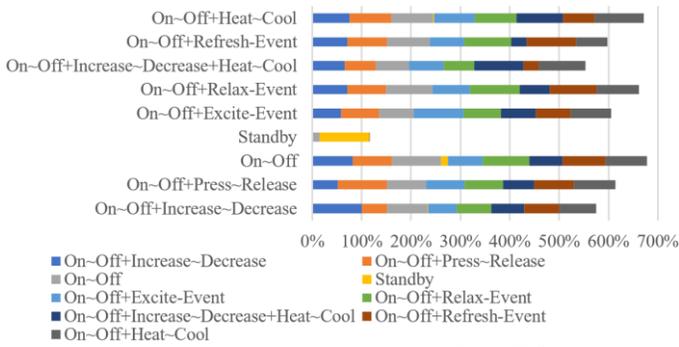

Fig. 13. Cross-similarities for different events using a Luyao-LY-518A massager. ~ is the event transition symbol, and + is the event separation symbol.

Table III Self-similarities for Some Complex Appliances

| Appliance | Brand | Application | Event | Similarity |
|---|---|---|---|---|
| cell phone | iPhone-10 | AliPay | open~close | 93.34% |
| cell phone | iPhone-10 | call | hang-on~hang-off | 94.83% |
| cell phone | iPhone-10 | screen | on~off | 93.29% |
| cell phone | iPhone-10 | text message | send~receive | 91.21% |
| cell phone | iPhone-10 | WeChat | open~close | 92.99% |
| cell phone | iPhone-10 | WeChat | open-scan~close-scan | 94.97% |
| laptop | Dell-G3-3590 | Acrobat Reader | open~close | 94.25% |
| laptop | Dell-G3-3590 | Chrome | open~close | 92.08% |
| laptop | Dell-G3-3590 | PyCharm | open~close | 94.03% |
| laptop | Dell-G3-3590 | Teams | open~close | 96.59% |
| laptop | Dell-G3-3590 | Word | open-document~close-document | 95.90% |
| laptop | Dell-G3-3590 | Word | save-as-start~save-as-end | 90.06% |
| laptop | Dell-Inspiron-15-7547 | Chrome | open~close | 87.99% |
| laptop | Dell-Inspiron-15-7547 | Edge | open~close | 89.92% |
| laptop | Dell-Inspiron-15-7547 | Excel | open~close | 88.11% |
| laptop | Dell-Inspiron-15-7547 | PowerPoint | open~close | 82.50% |
| laptop | Dell-Inspiron-15-7547 | Word | open~close | 92.33% |
| laptop | MacBook-Air-A1466 | Chrome | open~close | 96.32% |
| laptop | MacBook-Air-A1466 | Excel | open~close | 96.83% |
| laptop | MacBook-Air-A1466 | Map | open~close | 97.63% |
| laptop | MacBook-Air-A1466 | OneNote | open~close | 98.15% |
| laptop | MacBook-Air-A1466 | Word | open~close | 96.12% |
| pad | Surface-1 | Camera | open~close | 90.62% |
| pad | Surface-1 | OneDrive | open~close | 83.24% |
| pad | Surface-1 | OneNote | open~close | 85.53% |
| pad | Surface-1 | PowerPoint | open~close | 91.96% |
| pad | Surface-1 | Word | open~close | 88.25% |

Notes: ~ is the event transition symbol.

Table IV Self-similarities for Some Simple Appliances

| Appliance | Brand | Event | Similarity |
|---|---|---|---|
| cell phone | iPhone-10 | power-on-charging-on~power-on-charging-off | 92.32% |
| eye massager | Desleep-de-f09 | power-off-charging-on~power-off-charging-off | 94.96% |
| fan | Midea-kyt2-25 | on~off+increase~decrease | 98.16% |
| fan | Midea-kyt2-25 | on~off+rotate~halt | 100.00% |
| fluorescent lamp | Osram-stl-t412w-03wt | on~off | 97.10% |
| hairdryer | Panasonic-eh-nd11 | on~off | 90.50% |
| hairdryer | Panasonic-eh-nd11 | on~off+increase~decrease | 94.92% |
| hairdryer | Philips-hp8120 | on~off | 93.41% |
| hairdryer | Philips-hp8120 | on~off+increase~decrease | 94.02% |
| hand warmer | Rainbow-dr30-1 | charging-on~charging-off | 96.97% |
| heater | Xianfeng-dyt-z2 | on~off | 97.58% |
| heater | Xianfeng-dyt-z2 | on~off+increase~decrease | 92.09% |
| incandescent lamp | Osram-stl-t412w-03wt | on~off | 95.53% |
| laptop | Dell-G3-3590 | power-off-start~power-off-end | 98.48% |
| laptop | Dell-G3-3590 | power-on-start~power-on-end | 86.10% |
| laptop | Dell-Inspiron-15-7547 | power-off-start~power-off-end | 96.58% |
| laptop | MacBook-Air-A1466 | power-off-charging-on~power-off-charging-off | 96.23% |
| laptop | MacBook-Air-A1466 | power-on-charging-on~power-on-charging-off | 96.88% |
| LED lamp | Dengbeiwang-2013-III | on~off | 91.31% |
| massager | Keboer-kb-609a | on~off | 93.34% |
| massager | Luyao-ly-518a | on~off+heat~cool | 99.70% |
| massager | Luyao-ly-518a | on~off+press~release | 98.48% |
| monitor | AOC-27b2h | on~off | 98.98% |
| mosquito repeller | Zhuangchen-scj-ic-169 | on~off | 93.37% |
| pad | HUAWEI-honor-x2-gem-703l | power-off-charging-on~power-off-charging-off | 97.95% |
| pad | Kindle-d01100 | power-off-charging-on~power-off-charging-off | 95.89% |
| pad | Surface-1 | power-on-charging-on~power-on-charging-off | 79.94% |
| presenter | Knorvay-n75c | power-off-charging-on~power-off-charging-off | 98.23% |
| razor | Philips-rq310 | power-off-charging-on~power-off-charging-off | 93.67% |
| razor | Xiaoshi-f1-bk | power-off-charging-on~power-off-charging-off | 96.19% |
| router | Xiaomi-4a-1200m | on~off | 98.24% |
| toothbrush | Panasonic-doltz-ew-dm71 | charging-on~charging-off | 89.19% |
| toothbrush | Philips-sonicare-hx6530 | charging-on~charging-off | 74.45% |

Notes: ~ is the event transition symbol, and + is the event separation symbol.

## III. ENERGY DATA COLLECTION

This paper offers an OCR-based non-intrusive energy data collection method. The method is based on the proposed sub-sample convolutional neural network and an auto-correction mechanism. The proposed method is more convenient and less intrusive than traditional data collection approaches that require sophisticated network or hardware configurations.

### A. Overall Procedure

Appliance energy consumption data are firstly displayed on the monitor of a power meter. The camera then captures video frames of the meter and sends them to the server. The server is responsible for three consecutive tasks: 1) recognize the energy data in image format; 2) fine-tune the energy data according to physic rules; 3) store the energy data into the EAD dataset. Fig. 14 offers a graphical illustration of the energy data collection procedure.

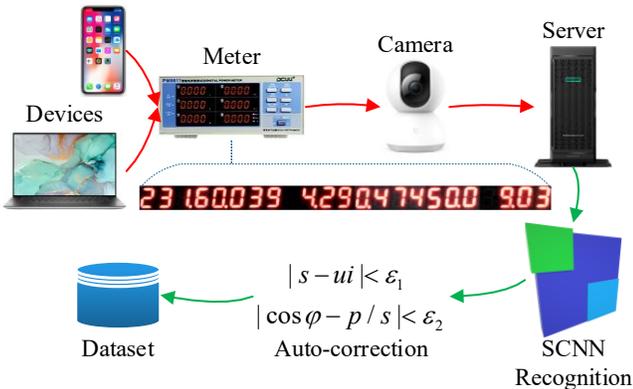

Fig. 14. Energy data displayed on the power meter are recognized, fine-tined, and stored in EAD. The red and green arrows represent the hardware and software processes of the energy data collection procedure.

Fig. 15 shows how a camera collects energy data from a power meter.

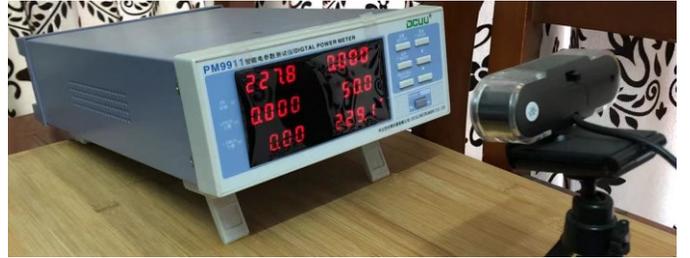

Fig. 15. Energy data are collected using a power meter and a webcam.

### B. Subsample Convolutional Neural Network

Inspired by the human retina structure that the fovea region has a higher resolution than its surroundings, an SCNN is proposed to extract additional features from a given sub-region of an input tensor, *e.g.,* the central area of an image can be a sub-region of the image. In this work, energy data from the power meter monitor are recognized by SCNN. The sub-region lies at the lower right corner of a digit where a **decimal point** exists. Given that a decimal point occupies a tiny area of a digit image, SCNN improves decimal point recognition accuracy because it processes important information emphasized by the sub-region. Fig. 16 shows the positions of the digit image sub-regions.

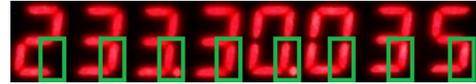

Fig. 16. The green bounding boxes are sub-regions of digit images.

To clarify illustrations, define $C$ as both the number of filters and the number of classification categories; $D$ as the height and width of a filter; channel count $\times$ height $\times$ width as the dimension order; $(\cdot)$ as the neural network layer index. The SCNN has three convolution layers, two fully connected layers, and a softmax regression layer. There are $C$ filters in the first convolution layer (Conv1), each of which has a dimension of $3 \times D \times D$. There are $C$ filters in the second convolution layer (Conv2), each of which has a dimension of $C \times D \times D$. Let $\mathbf{A}^{(2)}$ with a dimension of $C \times P \times Q$ be the output tensor of Conv2; $\mathbf{S}^{(2)}$ with a dimension of $C \times \lfloor \gamma_h P \rfloor \times \lfloor \gamma_w Q \rfloor$, $\gamma_h, \gamma_w \in (0,1)$ be the sub-tensor of $\mathbf{A}^{(2)}$ and also the input tensor for the subsample convolution layer (SConv), where $\lfloor x \rfloor$ is the largest integer less than $x$. The outputs of Conv2 and the SConv are flattened to a fully connected (FC) layer and a sub-fully connected (SFC) layer. Then FC and SFC are combined to form a softmax regression layer with $C$ outputs. Layer normalization [17] and leaky-relu activation function [18] are applied to all convolution layers. Fig. 17 illustrates the architecture of the subsample CNN.

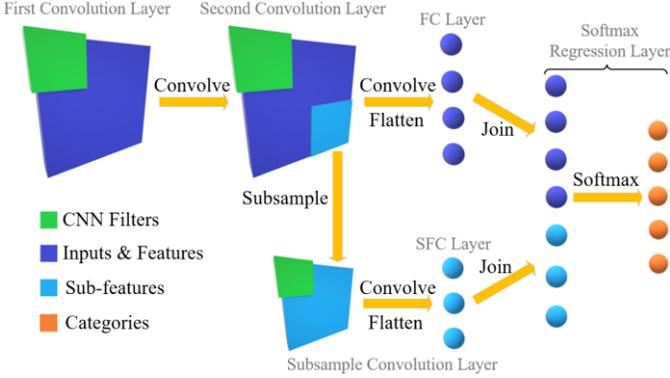

Fig. 17. SCNN improves recognition accuracy by processing information emphasized by the sub-region.

In the energy data collection scenario, hyperparameters settings are as follows: $C = 21$, where categories 1~10 represent digits 0~9, categories 11~20 represent digits with decimal points 0.~9., category 21 represents a blank area; $D = 3$; $\gamma_h = \gamma_w = 0.5$, where $\mathbf{S}^{(2)}$ is the sub-tensor on the lower right corner of $\mathbf{A}^{(2)}$. The height and width of the digit image are 57 and 42 pixels. A padding of 4 pixels is added to each digit image. When training SCNN, the learning rate is 0.01, 105,000 digit images are collected, 90% are used for training, and 10% are used for testing. The test set accuracy is 100%.

SCNN is proven to be more accurate and may converge faster than a conventional convolutional neural network (CNN). Let $a^{(2)} \in \mathbf{A}^{(2)}$ be a scalar from the subsampled region of $\mathbf{A}^{(2)}$. Since only reshaping occurs between Conv2 and FC, there exists a unique $x^{(FC)}$ scalar in FC such that $a^{(2)} = x^{(FC)}$. Likewise, since only subsampling occurs between Conv2 and SConv, there exists a unique $x^{(SConv)}$ scalar in SConv such that $a^{(2)} = x^{(SConv)}$. According to the chain rule, the partial derivative of $J$ with respect to (w.r.t.) $a^{(2)}$ is:

$$\frac{\partial J}{\partial a^{(2)}} = \frac{\partial J}{\partial x^{(FC)}} \frac{\partial x^{(FC)}}{\partial a^{(2)}} + \frac{\partial J}{\partial x^{(SConv)}} \frac{\partial x^{(SConv)}}{\partial a^{(2)}} = \frac{\partial J}{\partial x^{(FC)}} + \frac{\partial J}{\partial x^{(SConv)}} \quad (10)$$

When $\frac{\partial J}{\partial x^{(FC)}} \frac{\partial J}{\partial x^{(SConv)}} > 0$, a consensus is reached by FC and SConv regarding the output category. Thus, $\left| \partial J / \partial a^{(2)} \right|$ will be larger in SCNN than in CNN. The model will converge faster.

When $\frac{\partial J}{\partial x^{(FC)}} \frac{\partial J}{\partial x^{(SConv)}} < 0$, a discrepancy occurs between FC and Sconv regarding the output category. Therefore, $\left| \partial J / \partial a^{(2)} \right|$ will be smaller in SCNN than in CNN. The likelihood of wrong classifications will be lower.

When $\frac{\partial J}{\partial x^{(FC)}} \frac{\partial J}{\partial x^{(SConv)}} = 0$, either $\partial J / \partial a^{(2)}$ or $\partial J / \partial x^{(FC)}$ is $0$. Thus, vanishing gradient problems are less likely to happen, making the model easier to converge.

Although SCNN is only applied to energy data collection in this paper, SCNN can be applied to other computer vision tasks when sub-regions contain critical information.

### C. Auto-correction

An auto-correction mechanism is introduced to fix potential measurement errors of power meters by making physics quantities consistent. Let the energy vector be $\mathbf{e} = (u, i, s, p, \cos\varphi)$, where scalars in $\mathbf{e}$ denote voltage, current, apparent power, active power, and power factor. The energy vector should satisfy the following constraints, where $\varepsilon_1$ and $\varepsilon_2$ are maximum errors allowed:

$$\begin{aligned} |s - ui| &< \varepsilon_1 \\ |\cos\varphi - p/s| &< \varepsilon_2 \end{aligned} \quad (11)$$

If both constraints are satisfied, no correction is required. When only the first constraint is satisfied, $u$, $i$, and $s$ are correct, while at least one of $p$ and $\cos\varphi$ is incorrect. Thus, two energy vector candidates exist when $p$ or $\cos\varphi$ is assumed to be correct. $\cos\varphi$ and $p$ can be derived as:

$$\begin{aligned} \mathbf{e}_1 &= (u, i, s, p', \cos\varphi^*), \cos\varphi^* = p'/s \\ \mathbf{e}_2 &= (u, i, s, p^*, \cos\varphi'), p^* = s\cos\varphi' \end{aligned} \quad (12)$$

Where $x'$ is a value assumed to be correct, $x^*$ is a derived value. Likewise, when only the second constraint is satisfied:

$$\begin{aligned} \mathbf{e}_3 &= (u', i^*, s, p, \cos\varphi), i^* = s/u' \\ \mathbf{e}_4 &= (u^*, i', s, p, \cos\varphi), u^* = s/i' \end{aligned} \quad (13)$$

Similarly, when both constraints are incorrect, there are six energy vector candidates:

$$\begin{aligned} \mathbf{e}_5 &= (u', i', s^*, p', \cos\varphi^*), s^* = u'i', \cos\varphi^* = p'/s^* \\ \mathbf{e}_6 &= (u', i', s^*, p^*, \cos\varphi'), s^* = u'i', p^* = s^*\cos\varphi' \\ \mathbf{e}_6 &= (u', i^*, s', p', \cos\varphi^*), i^* = s'/u', \cos\varphi^* = p'/s' \\ \mathbf{e}_6 &= (u', i^*, s', p^*, \cos\varphi'), i^* = s'/u', p^* = s'\cos\varphi' \\ \mathbf{e}_6 &= (u^*, i', s', p', \cos\varphi^*), u^* = s'/i', \cos\varphi^* = p'/s' \\ \mathbf{e}_6 &= (u^*, i', s', p^*, \cos\varphi'), u^* = s'/i', p^* = s'\cos\varphi' \end{aligned} \quad (14)$$

The best energy vector candidate $\mathbf{e}_k$ is closest to the original energy vector $\mathbf{e}$:

$$\arg\min_k \|\mathbf{e}_k - \mathbf{e}\| \quad (15)$$

## IV. CONCLUSIONS

This paper offers a public energy activity dataset called EAD. EAD is designed for smart home energy research and has overcome various shortcomings of existing energy datasets, *e.g.*, insufficient labeling and lack of data type variety. All data points in EAD are fully labeled and have a high data diversity. Specifically, the appliance, brand, and event are labels of a simple appliance, whereas a complex appliance has an additional application/software label. EAD is designed to be the data source of DL-based energy projects so that homeowner energy consumption behaviors can be better understood. This paper provides graphs of sample data points as an overview of the dataset. Besides, energy dataset similarities are offered as a dataset quality indicator which is helpful for machine learning models. All energy data are collected using the non-intrusive OCR approach. The approach is based on SCNN and an auto-correction mechanism. In the future, EAD can be used for DL-based tasks such as energy-based cyber-attack detection and intelligent demand-side management.